# Intermediate State in Type I superconducting sphere: pinning and size effect.

I. Shapiro and B. Ya. Shapiro

Department of Physics and Institute of Superconductivity, Bar-Ilan University, 52900 Ramat-Gan, Israel

**Abstract**

Simulations, based on the time dependent Ginzburg-Landau equations show that the magnetization and spatial structure of the intermediate state are strongly affected both by the radius of the sphere and by concentration of the pinning centres. The intermediate state undergoes a transformation from a one-domain structure for small sphere to multi-domain structure in big spheres. In spheres where part of the superconducting material is replaced by 0.5% randomly distributed normal phase (dirty case), the intermediate state demonstrates a pronounced turbulence behaviour.

## I. Introduction

The subjects of a bulk superconductivity in a magnetic field is of many year efforts and well descried in numerous text books [1]. Long before BCS [2] the some very successful phenomenological theories of superconductors had been conceived. The most powerful tool was the Ginzburg-Landau (GL) theory, conceived in 1950 [3]. Considering superconductivity as a phase transition GL constructed gauge invariant Lagrangian theory. GL theory yields the Abrikosov parameter $\kappa = \lambda/\xi$ , where $\lambda$ and $\xi$ are the penetration magnetic field length and the coherence length correspondingly. In their original publication, Ginzburg and Landau showed that the solutions of their GL equations behave quite differently when $\kappa < 1/\sqrt{2}$ and $\kappa > 1/\sqrt{2}$ corresponding to what have come to be called a type-I and type-II superconductors. A cylinder a type-I superconductor infinite in the field direction expels the magnetic field from its interior for fields smaller than the thermodynamic critical field $H_c$ while for applied fields larger than the critical field, the sample is in the normal state, fully penetrated by the magnetic field. Magnetization in type-I superconductors demonstrates hysteresis at small $\kappa$ ($\kappa<0.42$) [4]. In a real system geometry external magnetic field reaches its critical value in some parts of the system while it still smaller in the rest (diamagnetic factor). This factor leads to the appearance of an intermediate state (IS), in which regions of both normal and Meissner state coexist. Due to the proximity effect these regions overlap blurring the borders between domains. Superconducting sphere with the diamagnetic factor (n=1/3) is the oldest studied example of such a system [5]. Experimental and theoretical effort is growing to obtain a general understanding of the problem [6–12]. The fundamental problem is that in a finite system, it is generally impossible to predict the topology of the intermediate state based solely on the energy minimization [7-19]. In fact in this approach we have to guess the topology of the IS and minimize then the energy which is not feasible for restricted geometry. From this point of view, exact numerical simulation of the time dependent GL equations in mesoscopic samples is the only way to study dynamics of the intermediate state from first principles with no initial assumption on the spatial distribution of the order parameter and magnetic field inside the sample. There are several factors that affect and often determine the topology of the intermediate state. Among them are sample size (size effect) and the inclusions of the normal phase (pinning centres) [20-25]. In spite of efforts allows this lines. First of all most of the numerical simulation were performed for large GL parameter (κ=0.4 for Pb in the Ref. [26]) so the dependence of the IS on κ was ignored along with the hysteresis behaviour of the magnetization. Secondly, the size effect and pinning effect affecting the IS structure have not been studied systematically. (In fact, in the overwhelming majority of published, papers flux pinning was omitted completely). Our calculation method, in contrast to others, allows learn of the dynamics of the IS inside the sphere. Here we study numerically the IS in two spheres "big" and "small" in a wide ranges of the magnetic fields at small GL parameter κ =0.18 (typical for Sn). We study the pinning effect on the IS and found a topologically induced change in a magnetization in the big spheres, where the IS adopts a multi-domain form. Hysteresis in magnetization recorded in the clean spheres was practically absent in dirty samples.

**II. Model and Numerical Simulations**

We consider here the magnetization as a function of the external magnetic field for type I superconducting spheres with radiuses in the range $\lambda \ll \xi_{eff} \ll R \ll \xi$, where $\lambda$ is the magnetic penetration depth, $\xi$ is the coherence length of the clean materials, while $\xi_{eff}^{-1} = \xi^{-1} + l^{-1}$ ($l$ is the mean free electron path) is the effective coherence length in dirty superconductors. The magnetic moment of a sphere with radius R subjected to an external magnetic field $H$ was studied by solving numerically the time dependent Ginzburg-Landau (GL) equations [27]. Starting from the dimensionless GL Hamiltonian [28]

$$G = \iiint d^3 r \left( -(1-T)|\psi|^2 + \frac{1}{2}|\psi|^4 + \left|\left(\frac{\partial}{\partial r_i} - -iA_i\right)\psi\right|^2 + \kappa^2(\partial \times A - H)^2 \right) \quad (1)$$

Here κ(T) is the Abrikosov parameter, ψ is the superconducting order parameter and $A$ is the vector potential in $\sqrt{2}\,\lambda H_c$ units, $H_c$ is the bulk thermodynamic magnetic field, the applied magnetic field H and the magnetic moment M is in $\sqrt{2}\,H_c$ units), while the coordinates are scaled by the coherence length [29]. The relaxation equations [30] have the form:

$$\frac{\partial \psi}{\partial t} = -\frac{\delta G}{\delta \psi^*} + f_\psi; \quad \frac{\partial A_\nu}{\partial t} = -\frac{1}{2}\frac{\delta G}{\delta A_\nu} + f_A \tag{2}$$

Where the $f_\psi$, $f_A$ are the magnitude of the order parameter and magnetic vector potential random noise. We use a link-variable approach and rectangular Cartesian grid (h is the step of the grid) **[31]** and the boundary conditions $H = H_z$ far from the sphere, and the normal order parameter gauge gradient vanishes at the sphere border. The set of equations (2) was solved numerically until a stationary state was reached. The stationary state solution provides the local magnetic field inside the sphere $H(r)$ and the magnetic induction $B = \int H(r)d^3r / \int d^3r$, while the magnetic moment has the usual form: $M(H) = (B - H)/4\pi$. The topology of the intermediate state and its characteristics strongly depend on the radius of the sphere where the order parameter inside the sphere was calculated for various magnetic field at the magnetization curve. We consider Abrikosov parameter κ=0.18 (in clean case) in two spherical samples: small, with radius $R = 6\xi$ (here ξ is the coherence length) and big, with $R = 15\xi$. (All calculations were performed for step h=0.25ξ and reduced temperature $\theta = \frac{T}{T_c} = 0.5$).

### III. Small Sphere *R=6 ξ, κ=0.18.*

Initial state of the superconducting sphere in zero magnetic field is plotted in Fig. 1 where colors scale denotes the magnitude of the superconducting electrons density changing from 0.5 (red color corresponds to the uniform superconducting state where $|\psi|^2 = 1 - \theta = 1/2$) to zero (black).

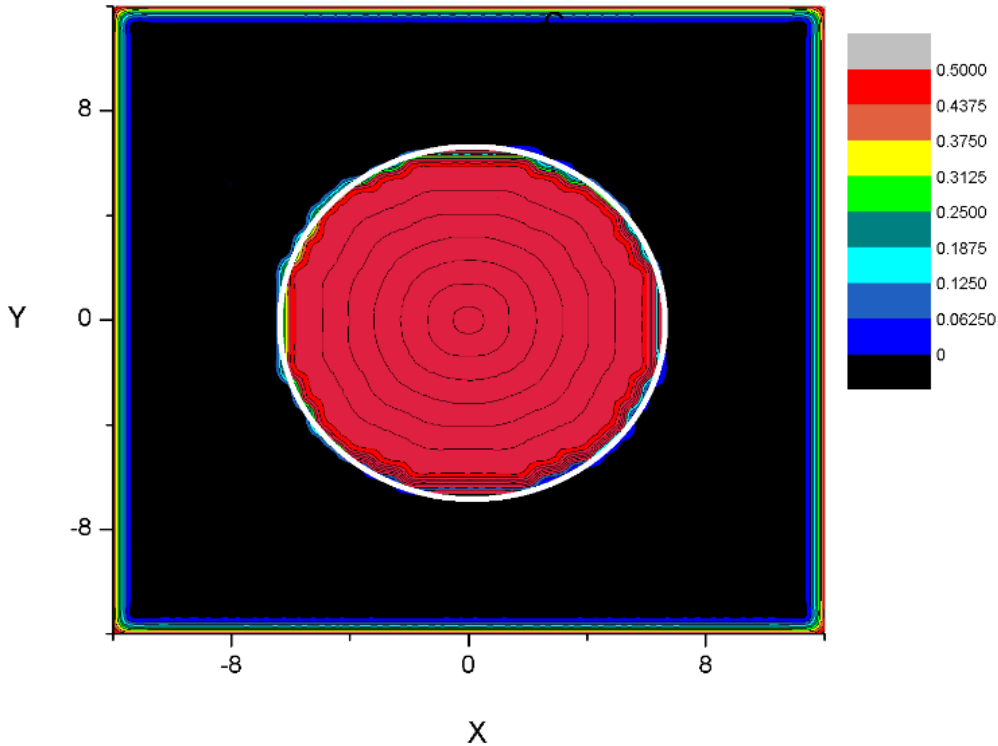

Fig.1. Spatial distribution of the superconducting electrons in zero magnetic field. The colour scale denoted the magnitude of the superconducting electrons density $|\psi|^2$. The magnetic field is directed in the z-direction.

The magnetization curve $M(H)$ was calculated both the external magnetic field increased from zero to critical field $H_{sh}$ denoted as forward direction (here $M(H_{sh}) = 0$) and in the backward direction, where the magnetic field decreases from $H_{sh}$ to zero. The lower, deeper, curve depicts the magnetization of the clean superconducting sphere while upper, more shallow curve shows the magnetization of the sphere with normal phase inclusions ("dirty case") amounting to 0.5% of the total volume of the sphere.

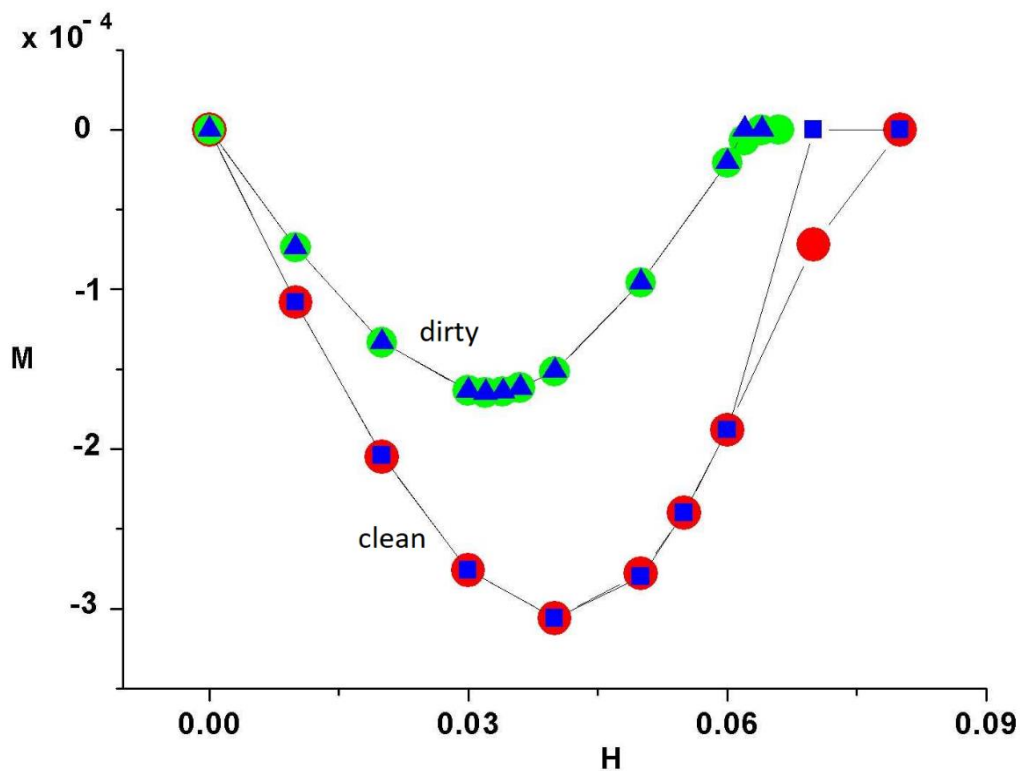

Fig.2 Forward and back magnetization of a small superconducting sphere as a function of the external magnetic field $H$.

This figure demonstrates weak hysteresis for the clean sphere superconductor (where the red circles for "forward" magnetization and blue square for backward magnetization) and for "dirty" sample. Green circles for "forward" magnetization coincide with "back" magnetization (blue triangles). On the other hand hysteresis is completely absent. The magnetization curves for a small sphere demonstrates reversibility at most values of the magnetic field values. The intermediate state inside the sphere shows strong difference between the clean and dirty samples as is shown in Fig.3 where density of the superconducting electrons $n_s = |\psi|^2$ is plotted for the same magnitudes of the magnetic fields in both cases in x-y and y-z projections.

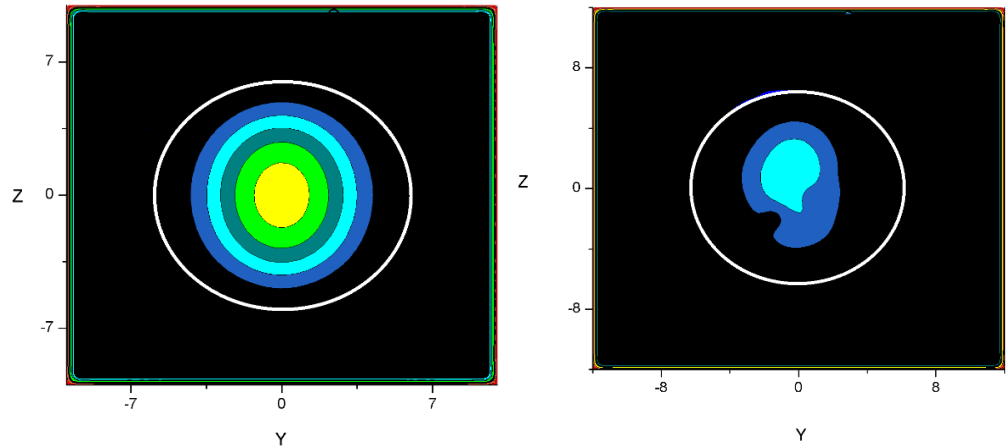

Fig.3. Density of the superconducting electrons inside small sphere in Y-Z projection for clean and "dirty" superconducting spheres. Magnetic field is 0.05.

Intermediate state in this case has an one domain structure with domain shape strongly dependent even on small concentration of the normal inclusions. This tendency is well pronounced at external magnetic field close to the critical field $H_s$ (see Fig.4). In clean sphere the superconducting domain stretches along the magnetic field becoming narrower in perpendicular direction while in dirty superconducting sphere superconductivity disappears inside the sample.

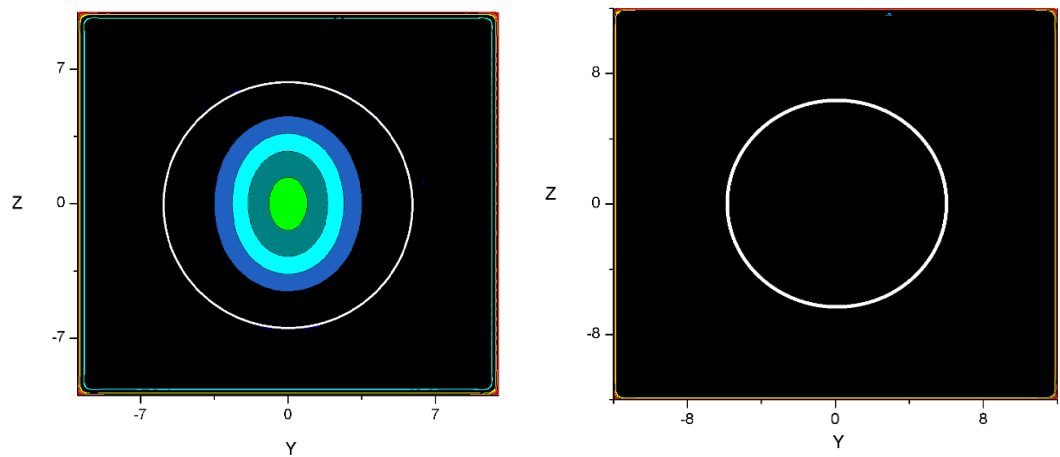

Fig. 4 Density of the superconducting electrons inside small sphere in Y-Z projection for clean (left) and "dirty" spheres. $R = 6\xi, \kappa = 0.18$. Magnetic field is 0.06.

It should be concluded that in a small clean sphere there is one-domain state while in dirty sphere the unformulated turbulent state appears.

## IV. Big Sphere $R=15\xi$, $\kappa=0.18$

Magnetic moment of superconducting sphere in this case is more complicated. It demonstrates hysteretic behaviour typical for bulk type-I superconductors with small κ parameter [29]. Magnetic moment as a function on the external magnetic field was calculated in four different protocols: "forward" direction for a. clean and b. dirty spheres subjected to the external magnetic field increases from zero to the critical field $H_{sh}$ and c. for magnetic field decreases from $H_{sh}$ to zero for clean and for dirty samples ("back" direction). Results are presented in Fig.5.

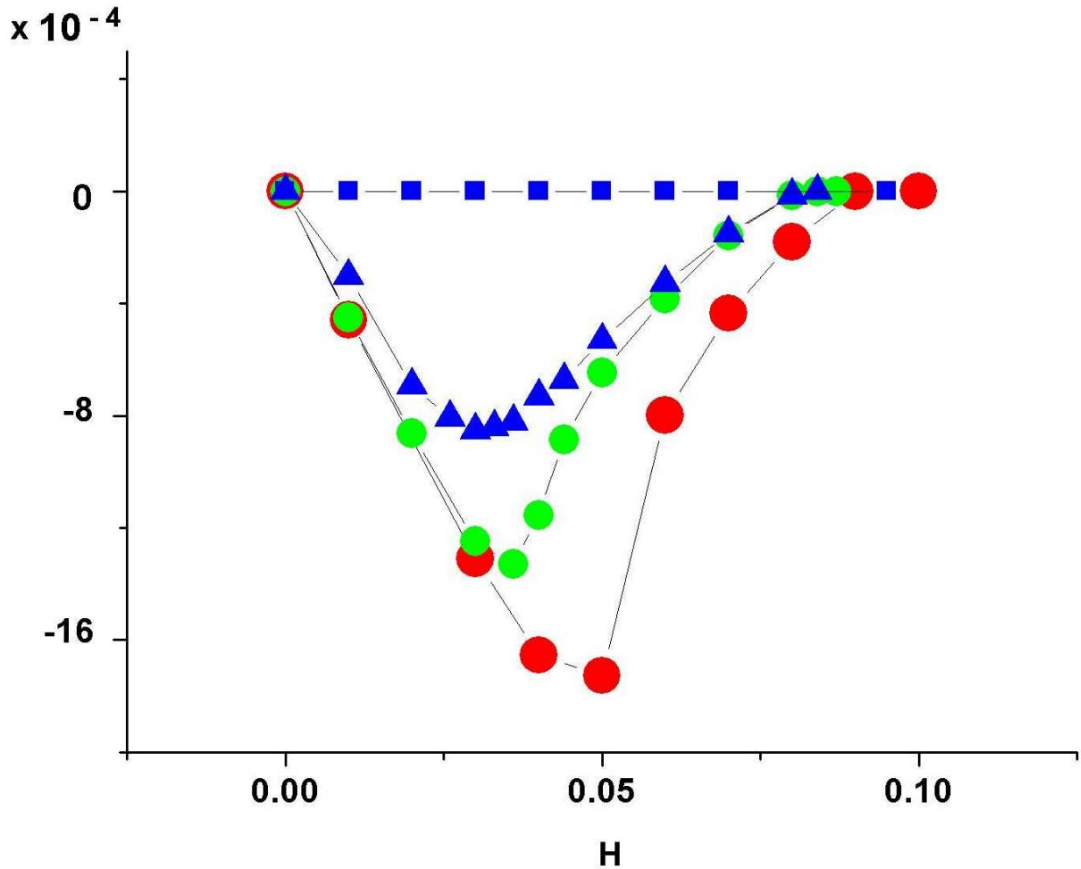

Fig.5. Magnetic moment of the superconducting sphere with radius $R = 15\xi$ and κ=0.18. Magnetizations for "forward" regimes are presented by red (clean) and green circles (dirty) while blue squares and triangles represent back direction in clean and dirty cases correspondingly.

The magnetization demonstrates both Meissner behaviour at small magnetic field and a well pronounced intermediate state evidenced by the long tail that extends to $H_{sh}$ where superconducting state is suppressed by the applied magnetic field and $M(H_{sh}) = 0$. Magnetization of the clean spheres demonstrates the well pronounced hysteresis typical for bulk type I superconductors where the metastable normal state appears in magnetic fields in the range $H_{sc} < H < H_{sh}$ where $H_{sc}$ is the surface critical magnetic field (here $H_{sc} = 0.69 H_{c2}$ and $H_{c2}$ is the second critical magnetic field). In our case, however, the unstable, hysteretic region extends down to zero magnetic field $(H_{sc} \to 0)$. The intermediate state in big spheres is complicated and topologically diverse. It contains the multi-domain superconducting to normal structure where the proximity effect smooths the boundaries between domains typical for macroscopic type-I superconductors while the density of superconducting electrons is spatially modulated and there is no sharp border between the domains. Typical spatial structures of the intermediate state domains are presented in Figs.6,7

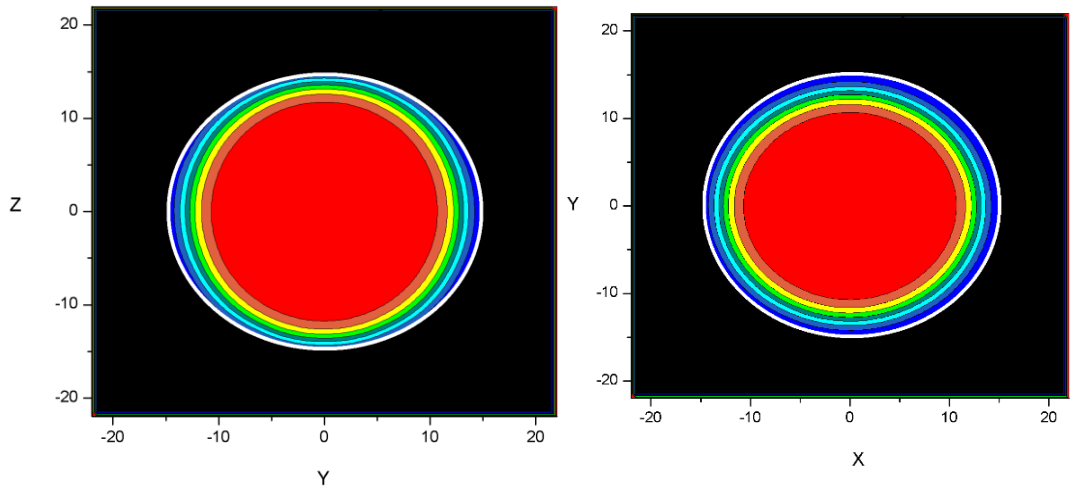

Fig.6 The intermediate state in the big superconducting sphere ($R = 15\xi, \kappa = 0.18$) for external magnetic field $H$=0.04 in clean sphere where Meissner state is well pronounced.

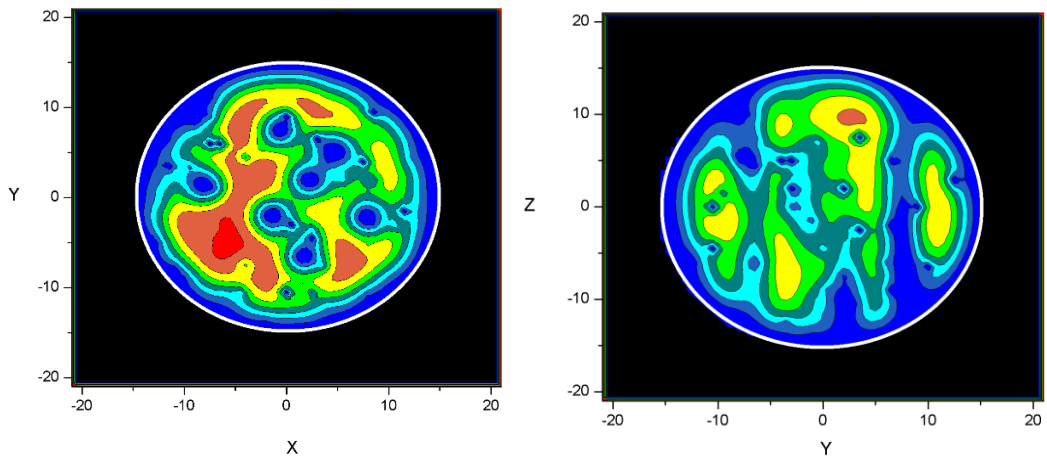

Fig.7 The intermediate state in the big superconducting sphere ($R = 15\xi, \kappa = 0.18$) for the external magnetic field *H*=0.04 in a dirty sphere where flux turbulence is exhibited.

In the external magnetic field *H*=0.04 the intermediate state in big clean sphere has a typical for the Meissner state shape while in dirty case, the IS manifests turbulent behaviour. In a larger magnetic field, $H = 0.06,$ the Meissner state in a clean sphere is broken and the intermediate state contains a set of domains (tubes) separated by normal regions (where weak superconductivity is induced by the proximity effect). In dirty sphere with randomly inclusions of the normal phase, the turbulent state becomes more pronounced and normal state percolation domains cross the sample (Fig.8,9)

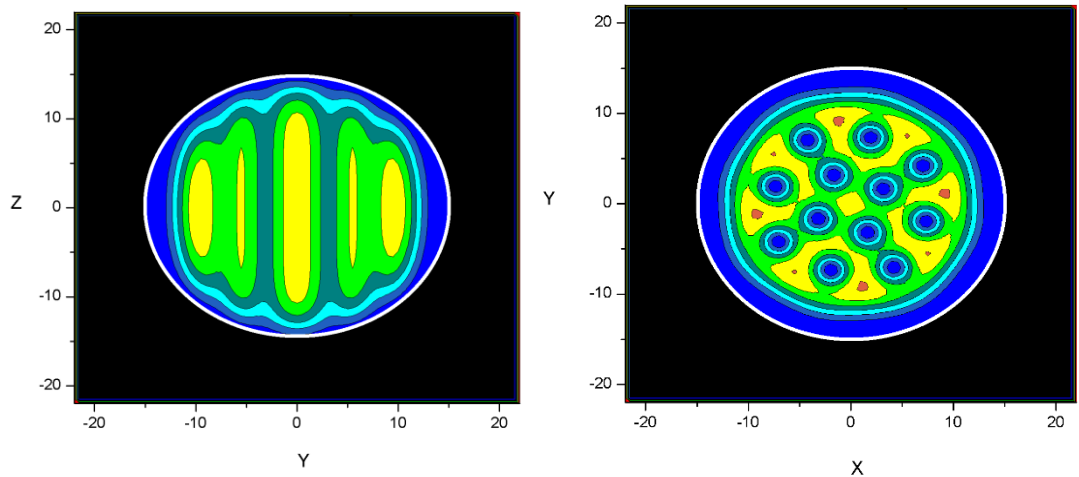

Fig.8 The intermediate state in a big clean superconducting sphere ($R = 15\xi, \kappa = 0.18$) in external magnetic field *H*=0.06.

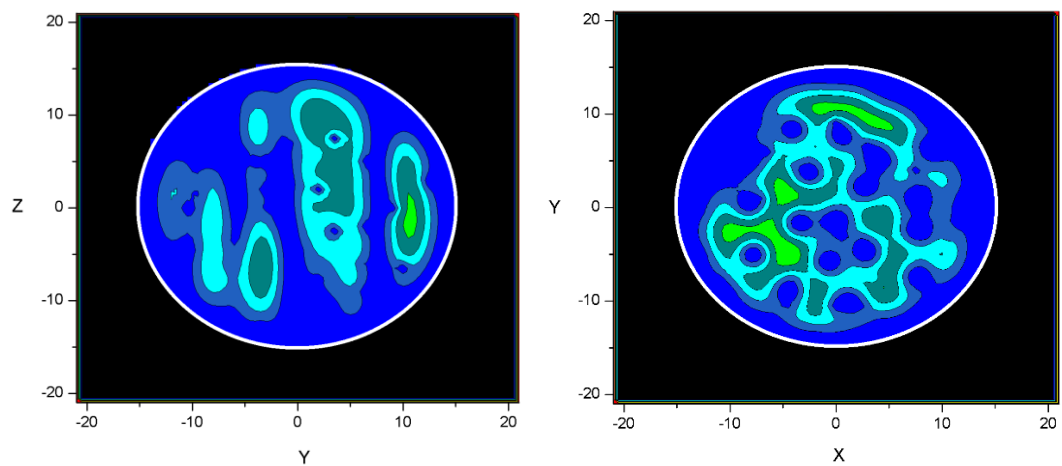

Fig. 9 The intermediate state in big dirty superconducting sphere ($R = 15\xi, \kappa = 0.18$) in external magnetic field H=0.06.

For $H = 0.06$ the Meissner intermediate state in clean big spheres is split by several domains (Fig.8) while in the dirty samples the turbulence state becomes more pronounced (Fig.9). The domains in clean sphere forms Abrikosov lattice similar to those in type-II superconductors. At magnetic fields $H = 0.07, H = 0.08$ (the critical magnetic field $H_{sh} = 0.1$ ) the amplitude of the domains decreases dramatically while the shape of the intermediate state and number of domains remains enhanced. The turbulent intermediate state in this case completely disappears while the sphere undergoes a transition to the normal state (Fig, 10,11.).

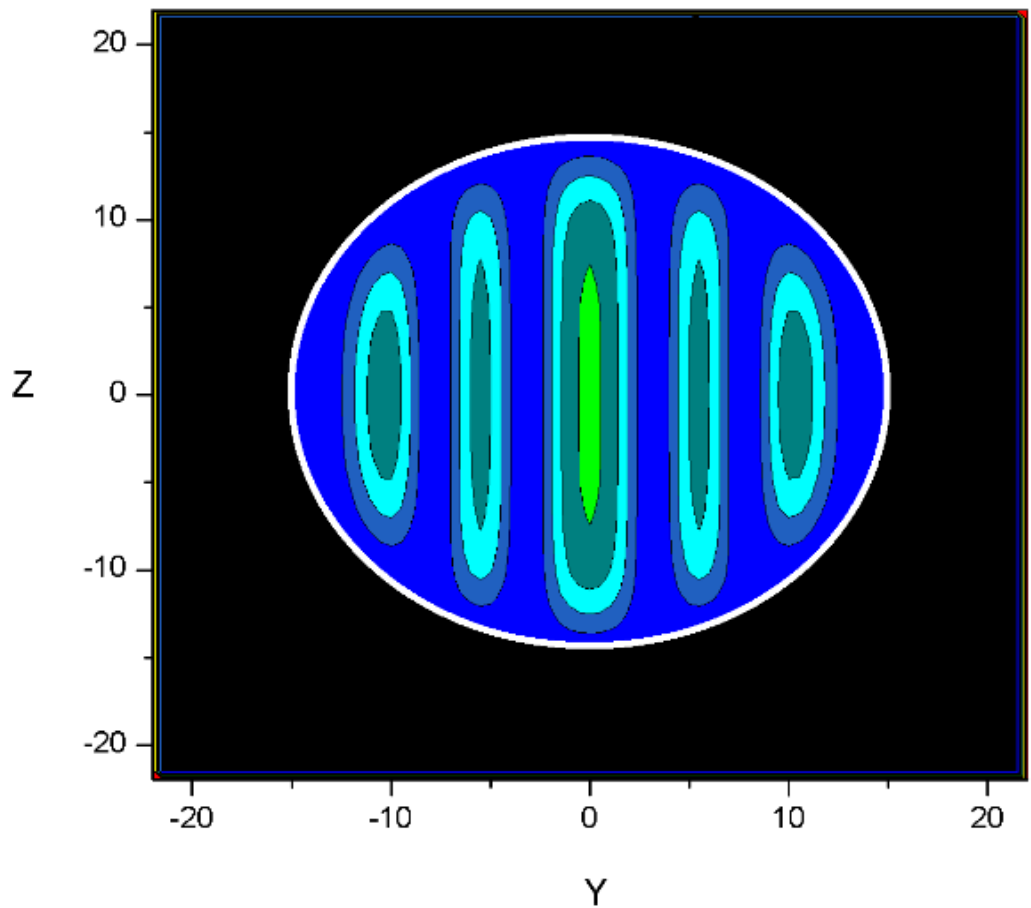

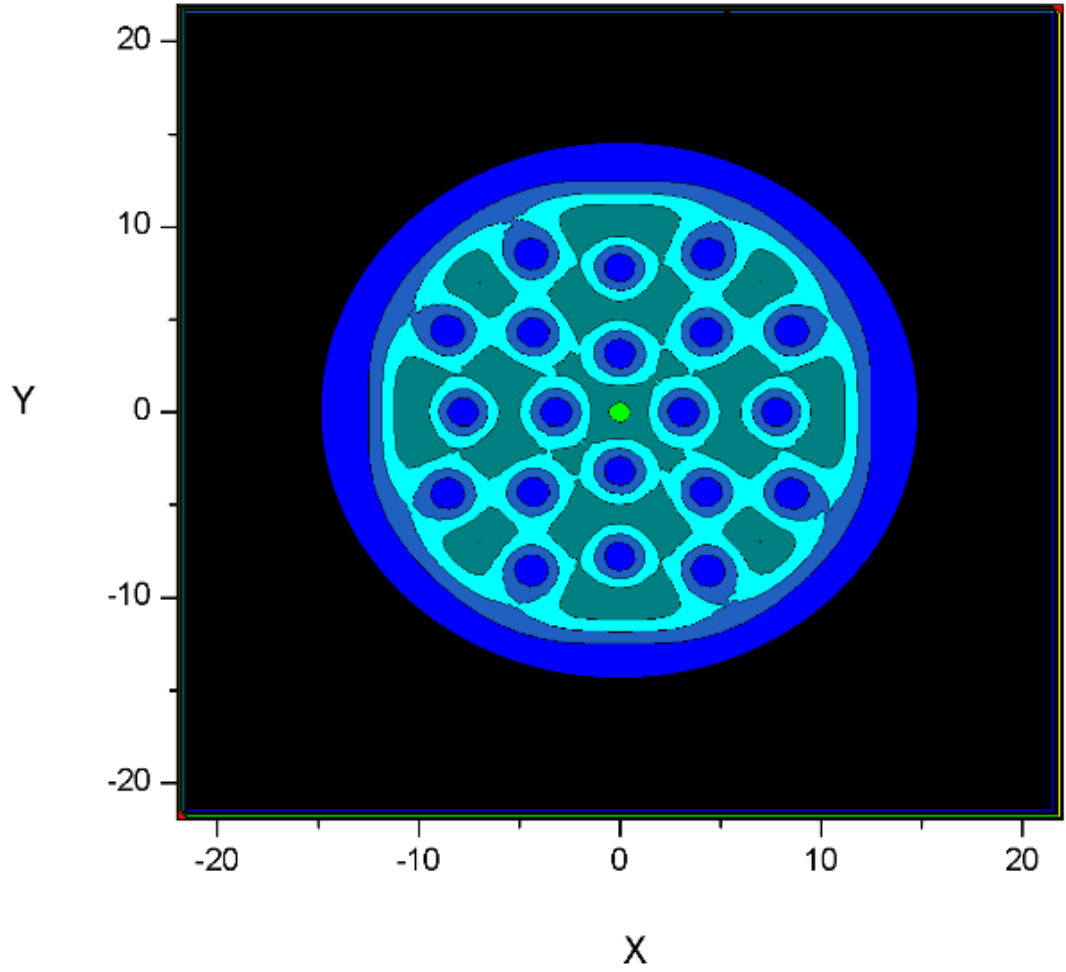

Fig.10. Density of superconducting electrons inside big clean sphere (*R=15, k=0.18,T = 0.5T<sub>c</sub>*) in the intermediate state for applied magnetic fields *H=0.07*.

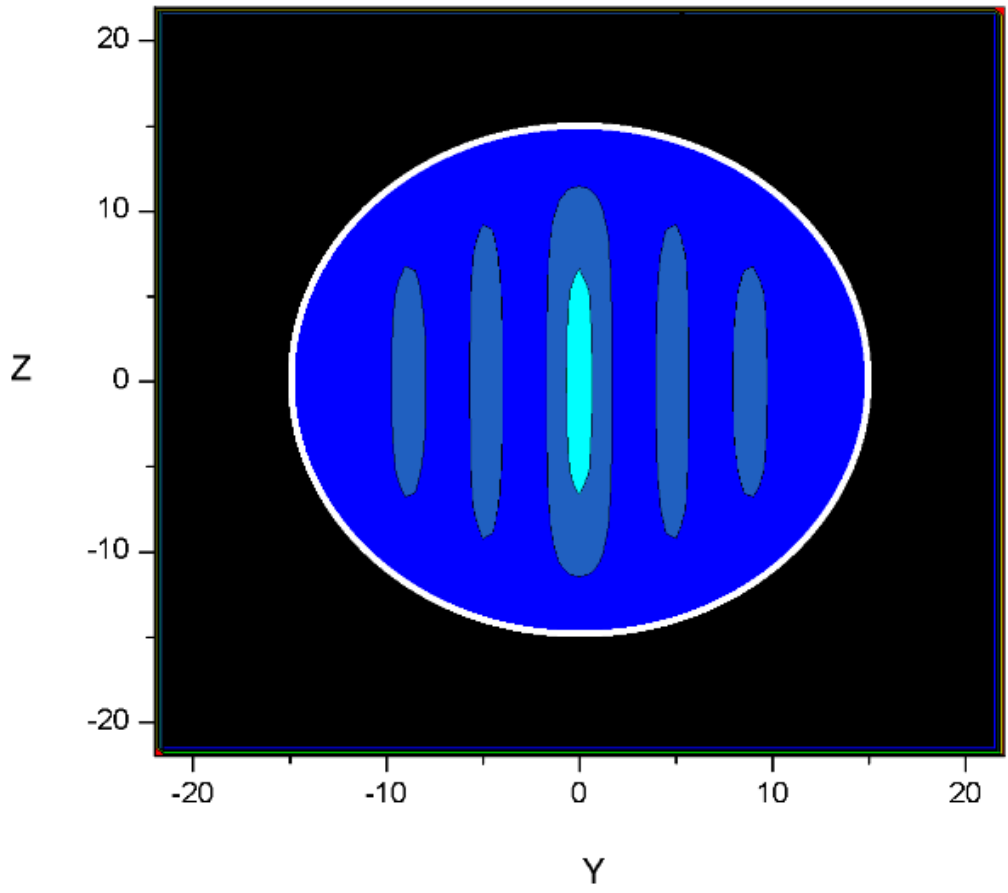

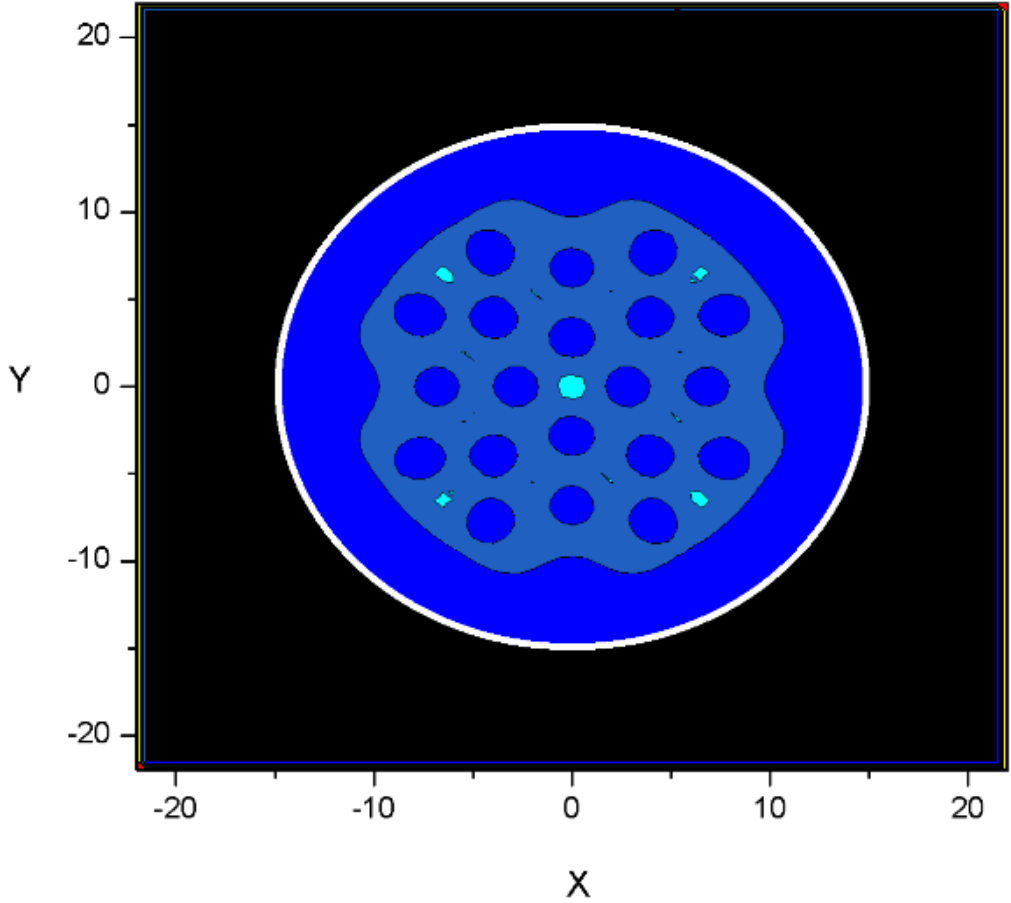

Fig.11. Density of superconducting electrons inside big clean sphere (R=15, k=0.18, $T = 0.5T_c$) in the intermediate state for applied magnetic fields. H=0.08.

**Summary**

The magnetic moment and the intermediate state as a function of the external magnetic field for small ($R = 6\xi$) and big ($R = 15\xi$) type I superconducting spheres have been calculated numerically in the framework of the time dependent Ginzburg-Landau equations. Both clean and dirty spherical samples (with 0.5% of the superconducting material of the spheres was replaced by the normal phase (dirty case) playing the role of the pinning centres), were studied. It was shown that the magnetic moments of the small sphere ($\kappa$=0.18, $T = 0.5T_c$ ) does not show hysteresis and irreversibility (Fig.2). In a big sphere the magnetization of the clean sphere demonstrates behaviour similar to the infinite system with a well pronounced hysteresis. The unstable normal hysteretic line (blue squares in Fig. 5) extends down to zero field due to the small surface barrier. The intermediate state in spheres demonstrates both domains (tubes) in the clean case and turbulent behaviour in dirty samples at the same magnitude of the external magnetic fields. The size of the sphere plays an extremely important role strongly affecting the structure of the intermediate state. In particular the intermediate state in a small clean sphere consists of one superconducting domain. This domain stretches along the magnetic field while in the dirty sphere at the same magnetic field the intermediate state demonstrates onset of turbulent behaviour (see Figs 6-11). These results are in a good qualitative agreement with the experiment presented in Ref. [6].

Acknowledgements

We thank Y.Yeshurun and A. Shaulov for succusful discussions. The simulations were co-funded by the European Regional Development Fund and the Republic of Cyprus through the Research Promotion Foundation (Project Cy-Tera)

**References.**

[1] Saint-James D., Sarma G.and Thomas E.J., 1969 Type II Superconductivity, Pergamon Press,
[2].Bardeen J, Cooper L.N.and.Schriffer J.R, 1957 Phys. Rev. B **106** 1629.
[3] Ginzburg V.L., Landau L.D.1950 Zh. Eksp. Teor. Fiz. **20** 1064 [ English translation in L. Landau 1965 Collected papers, Oxford Pergamon Press, 546].
[4] De Gennes P.G., 1966 Superconductivity of Metals and Alloys, W.A. Benjamin, Inc. New York-Amsterdam,
[5]Casimir H.B.G 1943 Physica **10** 757.
[6] Prozorov R. 2007  Phys. Rev. Lett. **98**  257001.
[7] Landau L.D. 1937 Sov. Phys. JETP **7**  371; Landau L. 1938 Nature (London) **141** 688; Choksi R., Kohn R.V., and Otto F   2004 J. Nonlinear Sci. **14**, 119 .
[8] Landau L.D., J. Phys. (Moscow) 1943 7 99.
[9] Livingston J.D. and DeSorbo W., 1969  Superconductivity, edited by Parks R.D. (Marcel Dekker, Inc., New York,) 2 1235.
[10] Huebener R.P. 2001 Magnetic Flux Structures of Superconductors (Springer-Verlag, New-York,).
[11]   Buckley K.B.W., Metlitski M.A. and Zhitnitsky A.R. 2004 Phys. Rev. Lett. 92 151102.
[12] Walgraef D., 1997 Spatio-Temporal Pattern Formation (Springer, Ney York,).
[13]  Gourdon C., Jeudy V and A. Cēbers, 2006 Phys. Rev. Lett. **96** 087002.
[14] Tinkham M. 2004 Introduction to Superconductivity, 2nd ed., Books on Physics, Dover,
[15] Menghini M. and Wijngaarden R.J. 2007 Phys. Rev. B **75**  014529.
[16] Bokil H. and Narayan O. 1997 Phys. Rev. B **56** 11195.
[17] Liu F., Mondelo M., and Goldenfeld N. 1991 Phys. Rev. Lett. **66**, 3071.
[18] Goldstein R.E., Jackson D.P., and Dorsey A.T., 1996 Rev. Lett. **76** 3818.
[19] Dorsey A.T. and Goldstein R.E., 1998 Phys. Rev. B **57** 3058.
[20]  Bose S. and  Ayyub P.  2014  Rep. Progr. Phys. **77** 116503 .
 [21] Das M. and  Wilson B.  2014 Adv. Nat. Sci. **6** 013001.
[22] Giaever I. and  Zeller H.  1968  Phys. Rev. Lett. **20**  1504.
[23] Reich S. ,  Leitus G. ,   Popovitz-Biro R.,  Schechter M. 2003  Phys. Rev. Lett. **91** 147001 .
[24] Hsu Y.J. ,   Lu S.Y. ,  Lin Y.F.  2006  Small **2**  268.
 [25] Shani L., Kumar V.B., Gedanken A., Shapiro I., Shapiro B.Ya., Shaulov A., Yeshurun Y.  2018 Physica C **546**  2.

[26] Muller A., Milosevic M. V., Dale S.E.C., Engbarth M. A., and Bending S. J., 2012 Phys. Rev. Lett., **109** 197003.
[27] Ginzburg V.L 1988 Non-equilibrium Superconductivity**, 174**, Nova Publishers
[28] Shapiro I., Pechenik E. and Shapiro B. Ya., 2001 Phys. Rev. B **63** 184520.
[29] Doria M.M. , Romaguera A.R.D.C. 2007 Peeters F. Phys. Rev. B **75** 064505 .
[30] Coskun E., Kwong M.K. 1997 Nonlinearity **10** 579 .
[31] Rothe H.J. 2005 "Lattice Gauge Theories", World Scientific, London